\documentclass[11pt,a4paper]{article}

\usepackage{fancyhdr}	
\usepackage{amsmath} \usepackage{amsfonts} \usepackage{amssymb}
\usepackage{wasysym}
\usepackage{xcolor}
\usepackage{graphicx}
\usepackage{multirow}
\usepackage[hang,footnotesize]{caption}

\renewcommand{\d}[1]{{\mathrm d}#1}
\newcommand{\fn}[2]{\mathinner{#1\mathopen{\left(#2\right)}}}

\newcommand{\mplank}[1]{{\tilde{M_\mathrm{p} ^{#1}}}}

\newcommand{\HE}[1]{H_{\mathrm{E}} ^{#1} }

\begin{document}

\title{The Possibility of Inflation in Asymptotically Safe Gravity}

\author{
Sungwook~E.~Hong${}^{1,2}$\thanks{eostm@muon.kaist.ac.kr},
Young Jae Lee${}^1$\thanks{noasac@muon.kaist.ac.kr}
and Heeseung~Zoe${}^{1,3}$\thanks{heezoe@gmail.com}
\medskip \\
\normalsize \\
\emph{ ${}^1$Department of Physics, KAIST} \\
\normalsize
\emph{Daejeon 305-701, Republic of Korea}
\\
\normalsize \\
\emph{ ${}^2$Department of Astronomy and Space Science,}\\
\normalsize
\emph{ Chungnam National University,}\\
\normalsize
\emph{ Daejeon 305-764, Republic of Korea}
\\
\normalsize \\
\emph{ ${}^3$Department of Physics,}\\
\normalsize
\emph{ Middle East Technical University,}\\
\normalsize
\emph{ 06531, Ankara, Turkey}
}

\maketitle

\abstract{
We examine the inflationary modes in the cubic curvature theories in the context of asymptotically safe gravity.
On the phase space of the Hubble parameter, there exists a critical point which corresponds to the slow-roll inflation in Einstein frame.
Most of the e-foldings are attained around the critical point for each inflationary trajectories.
If the coupling constants $g_i$ have the parametric relations generated as the power of the relative energy scale of inflation $H_0$ to the cutoff $\Lambda$, a successful inflation with more than 60 e-foldings occurs near the critical point.
}

\section{Introduction}

As is well-known, in four dimensional gravity, conventional wisdom gained in field theory does not work in reconciling unitary and renormalizability.
In particular, the unitarity of the non-renormalizable Einstein-Hilbert theory is ruined by the presence of a massive ghost when quadratic terms in the curvature---that make the theory renormalizable---are added \cite{stelle}.
This bleak state of affairs might change if Weinberg's long-standing conjecture of ``asymptotic safety'' works \cite{Weinberg}.
Leaving the details for an excellent review \cite{Niedermaier} and the references therein, let us note that asymptotic safety of gravity relies on the assumption that a non-gaussian fixed point exists for a finite gravity theory with infinitely many coupling constants and that the critical surface is finite dimensional, upgrading the theory to be a predictive one.
Recently, there has been a revival of interest in asymptotically safe gravity and are some encouraging results showing the possible existence of an asymptotically safe gravity \cite{Percacci}.

More recently, inflationary scenarios based on the asymptotically safe gravity are suggested \cite{Weinberginf}.
The effective action\footnote{We follow Weinberg's signature convention of the metric, $(+,-,-,-)$ and the Riemann tensor $R^{\mu}_{\nu \alpha \beta} = +\partial_{\alpha}\Gamma^{\mu}_{\nu \beta} + \cdots$.}
 is given by
\begin{multline}\label{zoewienberg}
I_{\Lambda}[g] = -\int \d{^4 x} \sqrt{-g} \left[ \Lambda^4 g_0(\Lambda) + \Lambda^2 g_1(\Lambda) R
+ g_{2}(\Lambda) R^2 \right. \\
\left. + g_{2a}(\Lambda) R^{\mu \nu} R_{\mu \nu} + g_{2b}(\Lambda) R^{\mu \nu \rho \sigma} R_{\mu \nu \rho \sigma} + \frac{g_{3}(\Lambda)}{\Lambda^{2}} R^3 \right. \\
\left. + \frac{g_{3a}(\Lambda)}{\Lambda^{2}} R R^{\mu \nu} R_{\mu \nu}
+ \frac{g_{3b}(\Lambda)}{\Lambda^{2}} R R^{\mu \nu \rho \sigma} R_{\mu \nu \rho \sigma}  + \ldots \right] \, ,
\end{multline}
where $\Lambda$ is the cutoff scale of the theory and $g_{i}(\Lambda)$'s are dimensionless coupling constants.
Weinberg showed that, without introducing a scalar field,
there exist de Sitter solutions and the modes of the Hubble parameter which can exit out of the pure de Sitter phase.

In this paper, building in Weinberg's work \cite{Weinberginf}, we explore the inflationary dynamics
by looking at the phase space of the Hubble parameter, ($H, \dot{H}$), determined by the classical equations derived from the variation of the action.
In the phase space, a proper trajectory for successful inflation should go to the non-inflationary era after giving more than 60 e-foldings.
We will see that the existence of such a trajectory depends on a certain algebraic equation describing the critical points 
where slow-roll inflation is possible. Moreover, this equation gives non-trivial conditions 
for the coupling constants and the relative energy scale of inflation to the cutoff.

The aim of this paper is to develop a method to analyze the Hubble parameter to see
how higher derivative terms would reproduce the standard inflationary history.
We are more interested in cosmology induced by $g_i$'s rather than the exact derivation of $g_i$'s.
There are some important comments as follows:
\begin{itemize}
\item We should be cautious on the ratio of inflationary energy scale $H_0$ and the cutoff scale $\Lambda$.
As noticed in \cite{Weinberginf, infASG, percacciinf, bonannocutoff},
the cutoff is to be optimally selected $\Lambda \sim \zeta H$ where $\zeta$ is a positive number of order unity.
Especially, in \cite{infASG, percacciinf, bonannocutoff}, the authors choose time dependence on cutoff as $\Lambda \sim H(t)$.
However, this could cause some problems in the context of standard inflationary cosmology.
Since a cosmological object specified by a certain comoving scale, $k$ at the present time should satisfy $k \lesssim  H_{\textrm{now}}$,
\begin{eqnarray}
\frac{1}{k} \gtrsim \frac{1}{H_{\textrm{now}}}
\end{eqnarray}
which means that all relevant length scales must be larger than the horizon size \cite{hindmarsh}.
Hence, one cannot discuss phenomena related to cosmological issues like power spectrum, and the inflationary cosmology is not sensible with the cutoff choice of $\Lambda \sim H(t)$.
Therefore, in this paper, we take the cutoff to be constant as in \cite{Weinberginf}.

\item We focus on the truncated actions to the cubic order, but there may be a danger that
the proper $g_i$'s for inflation may not fit with RG calculation of asymptotically safe gravity.
It is nontrivial to get a truncated action and to calculate RG flow equation of $g_i$'s \cite{reuterRG}.
In this work, however, our concern is not how to derive proper $g_i$'s for successful inflation in top-down
but whether inflationary possibility can give constraints on $g_i$'s in bottom-up.
\end{itemize}

This paper is organized as follows.
In Section~\ref{sec:einstein}, we consider certain conditions for successful inflation with enough e-foldings in the $R^2$- and $R^3$- gravities using their conformally related partners.
In Section~\ref{sec:hubble}, we consider the classical inflationary trajectories by analyzing the phase space of the Hubble parameter.
We discuss possible future works in Section~\ref{sec:discuss}.

\section{Conditions for slow-roll inflation}
\label{sec:einstein}
Before going into the general cases, we first discuss the possibility of inflation from a cubic curvature theory based on the Ricci scalar alone,
\begin{eqnarray}\label{eq:cubicaction}
I_{\Lambda}[g] = - \int \d^4 x \sqrt{-g}\left[ \Lambda^4 g_0 + \Lambda^2 g_1 R + g_{2}R^2
+ \Lambda^{-2} g_{3}R^3 \right].
\end{eqnarray}
We can investigate the possibility of inflation by directly tracing the Hubble parameter and find conditions on $g_i$'s that will produce enough e-foldings.
Also, we can rewrite this action as Einstein's gravity plus a scalar field by conformally rescaling the metric and discuss inflationary conditions.
So let us compare these two results to gain an understanding of how to deal with general higher derivative gravities.
In this section, the main goal will be to understand rather crudely the conditions on $g_i$'s which yield enough e-foldings,
namely, we will explore parametric relations between $g_i$'s.
In the following sections, accurate numerical simulations between $g_i$'s will be given.

\subsection{Classical analysis with the Hubble modes}

Now we will find the inflationary trajectories by directly analyzing the equations of motion coming from the variation of (\ref{eq:cubicaction})
as Weinberg did in \cite{Weinberginf}.
We will check whether it allows a de Sitter solution
and then look for the constraints on $g_i$'s by requiring enough e-foldings.

The physical degrees of freedom around FRW background, $ds^2 = dt^2 -a(t)^2 d\vec{x}^2$, are encoded in a single equation of the Hubble parameter, $H(t)= \dot{a}/a$,  which comes from the variation of the action \cite{Weinberginf, tye},
\begin{align}\label{eq:frN}
\mathcal{N}(t)  &\equiv -\frac{2}{\Lambda^4} \left( \frac{\delta I_{\Lambda}}{\delta g_{00}} \right)_\mathrm{FRW}  \\
\begin{split}
&=   -g_0  +g_1 \Lambda^{-2} \left( 6 H^2 \right)
- g_{2} \Lambda^{-4} (216H^2\dot{H} - 36 \dot{H}^2+72H \ddot{H}) \\
&\quad + g_{3} \Lambda^{-6} \left[ -864 H^6 +7776 H^4\dot{H} + 3240 H^2\dot{H}^2 -432 \dot{H}^3
+ 216H \ddot{H}(12H^2+6\dot{H}) \right] \\
&= 0 \, . \nonumber
\end{split}\end{align}
The dimensionless quantity, $H/\Lambda$, characterizes the energy scale of inflation to the cutoff scale.
This quantity is expected to be small, e.g. of the order of $10^{-5}$ in \cite{tye}, so that it would be helpful to collect the terms of (\ref{eq:frN}) accordingly, to analyze the behavior of the Hubble parameter.

\subsubsection{$g_3 \lesssim g_1 $}

We start considering a purely de Sitter solution by taking $H(t) = H_0$, and then (\ref{eq:frN}) yields
\begin{eqnarray}\label{eq:simplifiedfirst}
-g_0 + 6 g_1 \left( \frac{H_0}{\Lambda} \right)^2 -864 g_3 \left( \frac{H_0}{\Lambda} \right)^6 = 0 \, .
\end{eqnarray}
Since the third term is expected to be small, we can find a simple solution,
\begin{eqnarray}\label{eq:first}
H^2_0 = \frac{g_0}{6 g_1} \Lambda^2,
\end{eqnarray}
which relates $g_0$ and $g_1$ as
\begin{equation}
g_0 \thicksim g_1 \left( \frac{H_0}{\Lambda}\right)^2. \label{eq:g0g1}
\end{equation}

Now we investigate the classical behavior around the de Sitter solution in the linear approximation by putting a time dependent mode,
\begin{equation}
H(t) = H_0 + \delta(t).
\end{equation}
If $\delta(t) \propto \exp(\xi H_0 t )$ and $\xi >0$, then its e-folding number is $N_\mathrm{efold} \sim 1/\xi$ \cite{Weinberginf}.
By considering the lowest order terms in $H_0/\Lambda$ and $\delta(t)$ from (\ref{eq:frN}), one obtains
\begin{equation}
12 \frac{g_1}{\Lambda^2} \delta(t) - 216 \frac{g_{2}}{\Lambda^4} H_0 \dot{\delta}(t) = 0,
\end{equation}
whose solution is
\begin{equation}
\delta(t) \propto \exp \left[ \frac{g_1}{18 g_{2} } \left( \frac{\Lambda}{H_0} \right)^2 H_0 t \right].
\end{equation}
A discussion of the signature of the coupling constants is in order.
Attractive gravity requires $g_1 > 0$ and unitarity of the theory at the linearized level requires $g_2 >0$ for de Sitter and flat background
\footnote{Without referring to the unitarity issues, we might require $g_2 >0$ just to have an exit from de Sitter phase.}.
The requirement that one has more than 60 e-foldings,
\begin{equation}\label{eq:efold1}
N_{\textrm{efold}} = \frac{18 g_{2} }{g_1}  \left( \frac{H_0}{\Lambda} \right)^2  \gtrsim 60
\end{equation}
leads to
\begin{equation}
g_1 \thicksim g_2 \left( \frac{H_0}{\Lambda} \right)^2, \label{eq:g1g2}
\end{equation}
which is a similar relation as (\ref{eq:g0g1}).

\subsubsection{$g_3 \gg g_1$}

We have seen that $g_3$ term could not contribute to the construction of de Sitter phase.
However, if $g_3$ is large enough to compensate the suppression of $(H_0/\Lambda)^4$,
then the cubic terms would be important.
From (\ref{eq:simplifiedfirst}), we can read off the magnitude of $g_3$ for such a realization,
\begin{equation}
g_3 \thicksim g_1 \left( \frac{\Lambda}{H_0} \right)^4. \label{eq:g3g1}
\end{equation}
Then by considering higher order terms in $H_0/\Lambda$ and $\delta(t)$ from (\ref{eq:frN}), one has
\begin{multline}\label{eq:pert}
-5184\frac{g_3}{\Lambda^6}   H^5_0 \delta(t) +  12\frac{g_1}{\Lambda^2} H_0 \delta(t)
+ 7776 \frac{g_{3}}{\Lambda^6} H^4_0 \dot{\delta}(t)
- 216 \frac{g_{2}}{\Lambda^4} H^2_0 \dot{\delta}(t)  \\
+ 2592 \frac{g_{3}}{\Lambda^6} H^3_0 \ddot{\delta}(t)
- 72 \frac{g_{2}}{\Lambda^4} H_0 \ddot{\delta}(t) = 0.
\end{multline}
If we assume that $\delta(t)$ can give enough e-foldings, then we may put $\delta(t) \propto \exp \left( H_0/60 t \right) $ and
(\ref{eq:pert}) gives a solution for $g_3$ in terms of $g_1$ and $g_2$,
\begin{equation}
g_3 = \frac{50}{21057} g_1 \left(  \frac{\Lambda}{H_0}\right)^4 - \frac{181}{252684} g_2 \left(  \frac{\Lambda}{H_0}\right)^2 \, ,
\end{equation}
which implies
\begin{equation}
g_3 \sim 10^{-2 \sim 3} \times g_1 \left(  \frac{\Lambda}{H_0}\right)^4 \sim 10^{-2 \sim 3 } \times g_2 \left(  \frac{\Lambda}{H_0}\right)^2 \, ,
\end{equation}
and $g_3$ may be negative.

We have explored the relations between the coupling constants, $g_i$, in a toy model of $\fn{f}{R}$ gravity up to the cubic order.
If we rescale $g_i$ in compact forms,
\begin{equation}\label{eq:rescale}
\bar{g}_{0}   \equiv  {g}_{0} \left( \frac{\Lambda}{H_0}\right)^{6}, \,
\bar{g}_{1}   \equiv  {g}_{1} \left( \frac{\Lambda}{H_0}\right)^{4}, \,
\bar{g}_{2}   \equiv  {g}_{2} \left( \frac{\Lambda}{H_0}\right)^{2},
\end{equation}
then the condition for successful inflation with more than 60 e-foldings is
\begin{equation}\label{eq:inflcond}
\bar{g}_0 \sim \bar{g}_1 \sim \bar{g}_2,~~\textrm{and}~~~g_3 \lesssim 10^{-2 \sim 3 } \times \bar{g}_2.
\end{equation}
It is quite a crude estimation but provides a sound ground for numerical simulations.

\subsection{Conformally related Einstein action}

It is well known that $\fn{f}{R}$ gravity can be mapped to Einstein's gravity plus a scalar field after the metric is conformally scaled ($g_{\mu \nu} \rightarrow \Omega^2 g_{\mu \nu}$).
Here, we will conformally transform (\ref{eq:cubicaction}) into the Einstein action, and check its slow-roll inflationary condition.
(\ref{eq:cubicaction}) can be expressed as
\begin{equation}\label{eq:recubicaction}
I_{\Lambda}[g]  = - \Lambda^2g_1  \int \d{^4 x} \sqrt{-g} \fn{f}{R} \, ,
\end{equation}
where
\begin{equation}
\fn{f}{R} \equiv G_0 + G_1 R + G_2 R^2 + G_3 R^3   \, ,
\end{equation}
and
\begin{equation}
G_0 \equiv \Lambda^2 \frac{g_0}{g_1} \, , \,  G_1 \equiv 1 \, , \,
G_2 \equiv \Lambda^{-2} \frac{g_2}{g_1} \, , \, G_3 \equiv \Lambda^{-4} \frac{g_3}{g_1} \, .
\end{equation}
We assume $g_i$'s might have renormalization group flows and inflation would occur at a certain high energy scale.
It means that Planck mass $\mplank{}$ can be different from $10^{19}$GeV of our present low energy universe.
We have
\begin{equation}
\Lambda^2 g_1= \frac{1}{16\pi \tilde{G}} = \frac{\mplank{2}}{2} \, .
\end{equation}
According to (7.7) and (7.10) in \cite{perturbations},
the scalar field and its potential are given by
\begin{equation}\label{eq:phi_def}
\phi = \frac{\sqrt{6} \mplank{}}{2} \ln \frac{\partial f}{\partial R}
= \sqrt{3 \Lambda^2 g_1} \fn{\ln}{ G_1 + 2 G_2 R + 3 G_3 R^2} \, ,
\end{equation}
and
\begin{equation}\label{eq:V def}
\fn{V}{\phi} = \frac{\mplank{2}}{2} \frac{\fn{f}{R} - R {\partial f}/{\partial R}}{({\partial f}/{\partial R})^2 }
= \Lambda^2 g_1 \frac{G_0 - G_2 R^2 - 2 G_3 R^3}{(G_1 + 2 G_2 R + 3 G_3 R^2)^2} \, .
\end{equation}
Applying the slow-roll approximation, the equations of motion for the scalar field are
\begin{equation}\label{eq:slow_roll_one}
3 \HE{} \dot{\phi} = -\frac{\partial {V}}{\partial {\phi}} \quad\mbox{and}\quad \HE{2} = \frac{1}{3 \mplank{2}} \fn{V}{\phi} \, ,
\end{equation}
where the slow-roll parameters satisfy
\begin{equation}
\epsilon \equiv  \frac{\mplank{2}}{2} \left( \frac{\partial {V}/\partial {\phi}}{V} \right)^2 \ll 1 \quad\mbox{and}
\quad \eta \equiv  \mplank{2} \left( \frac{\partial^2 {V}/\partial {\phi^2}}{V} \right) \ll 1 \, .
\end{equation}
Let's assume that the slow-roll conditions are satisfied near a certain $\phi_0$ (and the corresponding $R_0$).
Then from $\epsilon \ll 1$, one has
\begin{equation}
\fn{\epsilon}{\phi_0} = 3 \left( \frac{2 G_0 + G_1 R_0 - G_3 R_0^3}{G_0 - G_2 R_0^2 - 2 G_3 R_0^3} \right)^2 \ll 1 \, ,
\end{equation}
which is sufficiently satisfied if one sets
\begin{equation}
2 G_0 + G_1 R_0 - G_3 R_0^3 \thickapprox 0 \, , \label{eq:epsilon}
\end{equation}
and the denominator should not vanish.
And from $\eta \ll 1$,
\begin{equation}\begin{split}
\fn{\eta}{\phi_0} &= \frac{(8 G_0 G_2 - G_1 ^2) + (24 G_0 G_3 + 2 G_1 G_2) R_0 + 12 G_1 G_3 R_0 ^2 + 2 G_2 G_3 R_0 ^3 - 3 G_3 ^2 R_0 ^4}
{3 (G_2 + 3 G_3 R_0 ) (G_0 - G_2 R_0 ^2 - 2 G_3 R_0 ^3)} \\
& \ll 1 \, ,
\end{split}\end{equation}
which is sufficiently satisfied if one sets
\begin{equation}\label{eq:preeta}
(8 G_0 G_2 - G_1 ^2) + (24 G_0 G_3 + 2 G_1 G_2) R_0 + 12 G_1 G_3 R_0 ^2 + 2 G_2 G_3 R_0 ^3 - 3 G_3 ^2 R_0 ^4 \thickapprox 0 \, ,
\end{equation}
and the denominator should not vanish.

Now, if we assume that $g_3 \thickapprox 0$, then (\ref{eq:epsilon}) and (\ref{eq:preeta}) give
\begin{equation}
R_0 = -\frac{2 G_0 }{G_1} \quad\mbox{and}\quad G^2_1 = 6 G_0 G_2,
\end{equation}
which implies
\begin{equation}
g_1^2 \sim g_0 g_2.
\end{equation}
In this case, the slow-roll conditions are satisfied if
\begin{equation}
g_0 \sim g_1 \left( \frac{H_0}{\Lambda} \right)^2 \quad\mbox{and}\quad
g_1 \sim g_2 \left( \frac{H_0}{\Lambda} \right)^2 \, .
\end{equation}

And if $g_3 \gtrsim 0$, (\ref{eq:epsilon}) and (\ref{eq:preeta}) give
\begin{equation}
(G_1 - 3 G_3 R_0 ^2)(G_1 + 2 G_2 R_0 + 3 G_3 R_0 ^2) \thickapprox 0 \, . \label{eq:eta}
\end{equation}
Since $G_1 + 2 G_2 R_0 + 3 G_3 R_0 ^2$ cannot be zero by (\ref{eq:phi_def}),
$G_1 - 3 G_3 R_0 ^2 \thickapprox 0$.
Then, from (\ref{eq:epsilon}),
\begin{equation}
R_0 = - \frac{3 G_0}{G_1} \quad\mbox{and}\quad
G_3 = \frac{G_1 ^3}{27 G_0 ^2} \, , \label{eq:G3}
\end{equation}
and its Hubble parameter satisfies
\begin{equation}
H_0 ^2 = \frac{1}{6} \frac{G_0 - G_2 R_0 ^2 - 2 G_3 R_0 ^2}{(G_1 +2 G_2 R_0 + 3 G_3 R_0 ^2)^2}
= \frac{1}{8} \frac{G_0}{G_1 ^2 - 3 G_0 G_2} = \frac{\Lambda^2}{8} \frac{g_0 g_1}{g_1 ^2 - 3 g_0 g_2} \, .
\end{equation}
This implies necessarily both $g_1 ^2$ and $3 g_0 g_2$ are in the same order as $g_0 g_1 (\Lambda / H_0)^2$;
that is,
\begin{equation}
g_0 \sim g_1 \left( \frac{H_0}{\Lambda} \right)^2 \quad\mbox{and}\quad
g_1 \sim g_2 \left( \frac{H_0}{\Lambda} \right)^2 \, ,
\end{equation}
and from (\ref{eq:G3}),
\begin{equation}
g_2 \sim 10^{-2} \times g_3 \left( \frac{H_0}{\Lambda} \right)^2 \, .
\end{equation}
These conditions are equivalent to what we found before.
Then we may claim that (\ref{eq:inflcond}) can necessarily satisfy the slow-roll conditions in the context of a scalar field plus Einstein's gravity.
Note that this exercise of mapping the $\fn{f}{R}$ action to Einstein's gravity plus a scalar field was necessary to see the slow-roll conditions that are well-defined for the scalar field.
Namely, in the pure higher derivative gravity, the slow-roll conditions were not transparent.

\section{Analysis on the phase space of the Hubble parameter}
\label{sec:hubble}

In the previous section, we considered the $\fn{f}{R}$ gravity case and found the conditions on $g_i$'s which give enough e-foldings.
Now we numerically study the classical inflationary trajectories and their e-foldings in the phase diagrams ($H, \dot{H}$) generated by the nonlinear equation of motion.

We will see that $g_i$'s affect the classical trajectories of the phase space and decide cosmological viability.
The trajectories in the phase space should satisfy the following conditions for successful inflation:
\begin{enumerate}
  \item The inflationary phase should be stopped leading to the standard post-inflationary history.
  A proper trajectory should go into the non-inflationary era, $\ddot{a}<0$.
  \item There must exist the attractor behavior for the naturalness of inflation.
  There should be some region in the phase diagram where all nearby trajectories converges to the non-inflationary era.
  \item A proper trajectory should give e-foldings more than 60 while moving to the non-inflationary era.
\end{enumerate}

\subsection{$\fn{f}{R}$ gravity}
The governing equation (\ref{eq:frN}), which was derived from $\fn{f}{R}$ gravity action (\ref{eq:cubicaction}),
can be rewritten with the redefinition of the Hubble parameter and its derivatives as
\begin{align}\label{eq:frN_three}
\fn{\bar{\mathcal{N}}}{h, h', h''} &\equiv -\frac{2 \Lambda^2}{H_0 ^6} \left( \frac{\delta I_{\Lambda}}{\delta g_{00}}\right)_\mathrm{FRW} = 0 \nonumber \\
\begin{split}
& =  \left( -\bar{g}_0 + 6 \bar{g}_1 h^2 - 864 g_{3} h^6 \right) + \left(-216 \bar{g}_{2} h^2 + 7776 g_{3} h^4 \right) h' \\
& \quad + \left( 36 \bar{g}_{2} + 3240 g_{3} h^2 \right) h'^2 - 432 g_{3} h'^3 \\
& \quad + \left[ \left(-72 \bar{g}_{2} h + 2592 g_{3} h^3 \right) + 1296 g_{3} h h' \right] h'' \, ,
\end{split}\end{align}
where $h \equiv H / H_0$, $h' \equiv \d{h}/\d{\left( H_0 t \right)}$,
and $h'' \equiv \d{^2 h}/\d{\left(H_0 t\right)^2}$.

\begin{figure}[hbt]
\centering
\includegraphics[width=0.47\textwidth]{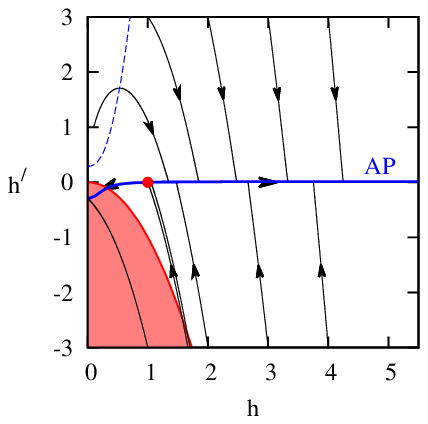}
\includegraphics[width=0.47\textwidth]{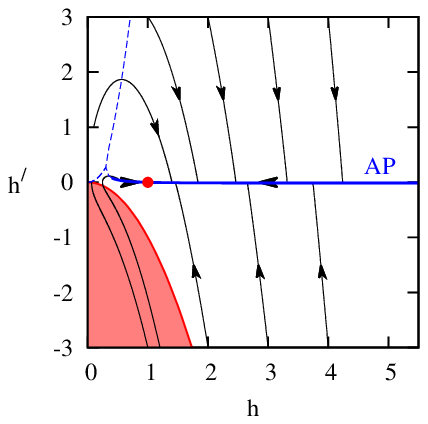}
\caption{
The general behavior of the classical trajectories in ($h, h'$) phase diagram with $g_3 = 0$.
Left: ($\bar{g_0}, \bar{g_1}, \bar{g_{2}},  g_{3}$) = (3, 0.5, 1, 0).
Right: ($\bar{g_0}, \bar{g_1}, \bar{g_{2}},  g_{3}$) = (3, 0.5, -1, 0).
Red area: $\ddot{a} < 0$, the non-inflationary era.
Red dot: $h_*$-point around which inflation is maximized, that is $h=1$ and $h' = h'' = 0$.
Blue dash: $h'' = 0$.
Blue solid: the asymptotic path(AP), that is $h''=0$, on which there exists the $h_*$-point.
In Left, the e-foldings, $N_\mathrm{efold}$ around the $h_*$-point is numerically estimated to be 200.
}\label{fig:good}
\end{figure}

Figure~\ref{fig:good} (Left) shows the general behavior of the classical trajectories in ($h, h'$) phase diagram with $g_3 = 0$.
There exists an asymptotic path(AP) with $h''=0$, where all classical trajectories converge to either $h = 0$ or $h = \infty$.
The de Sitter phase is at ($h,h'$) = (1,0), represented by a red dot in the figure.
A proper trajectory for successful inflation should pass through this dS point and end up with non-inflationary era, represented by the red area.

Since the AP satisfies $h'' = 0$ from (\ref{eq:frN_three}), along it, one must have
\begin{align}
\fn{\mathcal{B}}{h, h'} &\equiv -\fn{\bar{\mathcal{N}}}{h, h', 0} = 0 \nonumber \\
\begin{split}
&= 432 g_{3} h'^3 - (3240 g_{3} h^2 + 36 \bar{g}_{2}) h'^2 - (7776 g_{3} h^4 - 216 \bar{g}_{2} h^2) h' \\
& \quad + (864 g_{3} h^6 - 6 \bar{g}_1 h^2 + \bar{g_0}) = 0 \, .
\end{split}
\end{align}

Like a slow-roll point in inflationary model,
there may exist a point in the phase diagram that tends to keep its current state forever.
This is possible at certain points, $h=h_*$ where both $h'$ and $h''$ are zero, that is,
\begin{equation}\label{eq:fr_sepatrix}
\fn{\mathcal{C}}{h_*} \equiv \fn{\mathcal{B}}{h_*, 0} = 864 g_{3} h^6_*  - 6 \bar{g}_1 h^2_* + \bar{g}_0 = 0 \, .
\end{equation}
We call these points, i.e. the roots of (\ref{eq:fr_sepatrix}), $h_*$-points.
Actually, (\ref{eq:fr_sepatrix}) is a different expression of (\ref{eq:simplifiedfirst}).
If (\ref{eq:simplifiedfirst}) has a de Sitter solution, then (\ref{eq:fr_sepatrix}) would have $h_*=1$ as a root and the trajectories around this point in the phase diagram would be crucial for a successful inflation.

In Figure~\ref{fig:good} (Left),
$h_*$-point acts as an attractor along the normal direction to the AP and as a repeller along it.
If a trajectory approaches the AP at a point $h > h_*$
then it will follow the AP which goes far away from the non-inflationary region ($\ddot{a} < 0$) and the inflation never ends,
which does not fit to our universe.
On the other hand, if a trajectory approaches the AP at a point $h \ll h_*$, then it will soon fall into the non-inflationary region without generating enough e-foldings, which also does not fit to our universe.
Therefore, in order to explain our universe
the classical trajectory needs to approach the AP at the point where $h$ is equal to,
or slightly smaller than $h_*$.

However, $g_2 < 0$ severely changes the asymptotic behavior so that $h_*$-point is an attractor from all directions in Figure~\ref{fig:good} (Right).
In this case, if a trajectory approaches the $h_*$-point, then it would stay there forever and the inflationary phase cannot be stopped, which does not fit to our universe again.

We can easily see that for $g_3 =0 $, the sign of $g_2$ changes the asymptotic behavior around the $h_*$-point.
Around the $h_*$-point where $h=1$, since $h'^2 \ll h'$ and $h''=0$ in (\ref{eq:frN_three}), one has
\begin{equation}
\fn{\bar{\mathcal{N}}}{h, h', 0} \approx -\bar{g}_0 + 6 \bar{g}_1 h^2 -216 \bar{g}_{2} h^2 h' \approx 0,
\end{equation}
which gives
\begin{equation}\label{eq:easyg2}
h' \approx \frac{1}{\bar{g}_2} \left( \frac{\bar{g}_1}{36 } - \frac{\bar{g}_0}{216 h^2} \right) =
\frac{\bar{g}_1}{36 \bar{g}_2 h^2} \left( h^2-1 \right),
\end{equation}
from the fact that at $h=1$, one has $h'=0$, which implies $\bar{g}_0 = 6 \bar{g}_1$.
Now one can see that for $g_2>0$, $h'$ is a monotonically increasing function of $h$ around the $h_*$-point from (\ref{eq:easyg2}).
For $h<1$, one has $h'<0$ but for $h<1$, $h'>0$.
However, for $g_2<0$, $h'$ is a monotonically decreasing function of $h$ so that for $h<1$ one has $h'>0$ but for $h>1$,  $h'<0$.
This explains completely the asymptotic behavior around the $h_*$-points in Figure \ref{fig:good}.
Actually, $g_2<0$ is ruled out by the requirement of unitarity of the free theory, as we mentioned earlier, but now we check it is also not allowed in the context of inflationary cosmology.

\subsubsection{How to get an inflationary trajectory}
The fact that $h_*=1$ is a solution of (\ref{eq:fr_sepatrix})
would give tight constraints on $g_i$'s and be useful to predict the behavior along the AP of the phase diagram.
Putting $h_* =1$ in (\ref{eq:fr_sepatrix}) yields,
\begin{equation}\label{eq:g3}
 g_{3} = \frac{1}{864} \left( 6 \bar{g}_1 - \bar{g}_0 \right) \, .
\end{equation}
Let us note that we had four parameters ($g_0, g_1, g_2, g_3$) in (\ref{eq:simplifiedfirst}) to fix $H_0$ by solving (\ref{eq:simplifiedfirst}).
However, we have changed our parameter set to ($\bar{g}_0, \bar{g}_1, \bar{g}_2, H_0$) as (\ref{eq:fr_sepatrix}) and can fix $g_3$ by requiring $h_*=1$.

\begin{figure}[htb]
\centering
\includegraphics[width=0.47\textwidth]{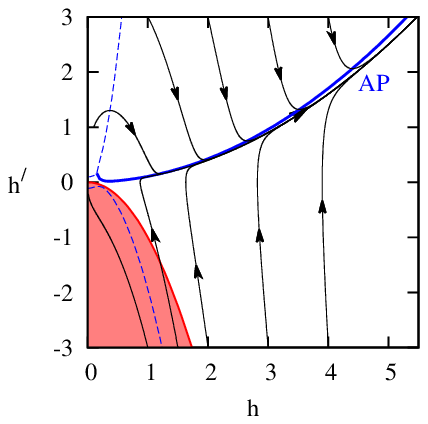}
\includegraphics[width=0.47\textwidth]{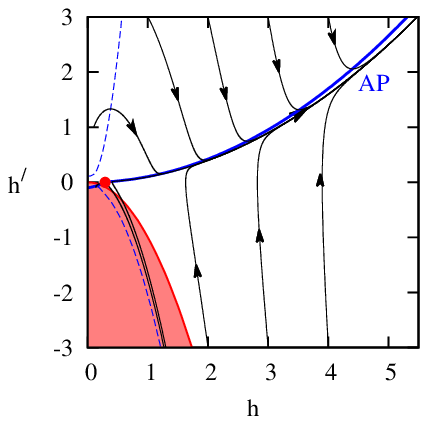}
\caption{
The general behavior of the classical trajectories in ($h, h'$) phase diagram when $g_i$'s do not satisfy (\ref{eq:g3}).
Left: ($\bar{g_0}, \bar{g_1}, \bar{g_{2}},  g_{3}$) = (1, 1, 1, 1).
Right: ($\bar{g_0}, \bar{g_1}, \bar{g_{2}},  g_{3}$) = (1, 1, 1, -1).
The asymptotic path(AP)s, blue solid lines, are the curves going to the top-right corner.
In Left, the $h_*$-point does not exist. In Right, $h_*$-point is at $h<1$.
In both cases, $N_\mathrm{efold} \ll 10$.}
\label{fig:example}
\end{figure}

Figure~\ref{fig:example} shows that there is no possible inflationary trajectory with enough e-foldings
when $g_i$'s do not satisfy (\ref{eq:g3}).
In this case, no $h_*$-point exists or even if it does exist, it is too close to the non-inflationary era to give enough e-foldings.

\begin{figure}[htb]
\centering
\includegraphics[height=0.33\textwidth]{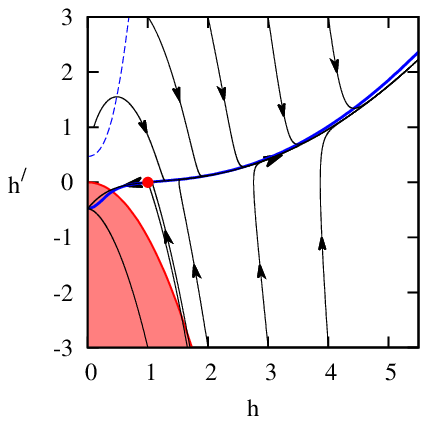}
\includegraphics[height=0.33\textwidth]{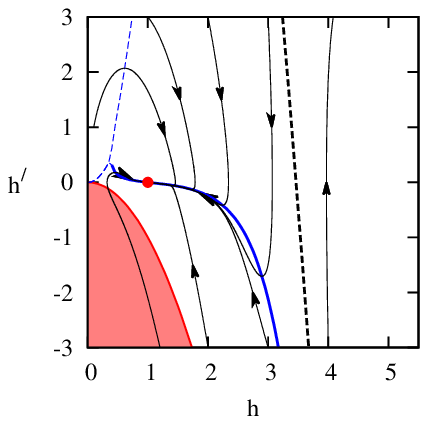}
\caption{
The general behavior of the classical trajectories in ($h, h'$) phase diagram with $g_3 < 0$.
Left: ($\bar{g_0}, \bar{g_1}, \bar{g_{2}},  g_{3}$) = (8, 1, 1, -1/432) with $N_\mathrm{efold} = 90$.
Right: ($\bar{g_0}, \bar{g_1}, \bar{g_{2}},  g_{3}$) = (8, 1, -1, -1/432).
The directions of trajectories around the divergence curve(black dash) given by (\ref{eq:h_diverge}) are opposite and the $h_*$-point is under it.
In Left, the divergence curve is too under the $h_*$-point so that it is not shown in the figure
and it does not affect the direction of trajectories near the $h_*$-point.
In Right, the divergence curve blows up and the direction of trajectories near the $h_*$-point is opposite to Left.
}
\label{fig:negativeg3}
\end{figure}

By using (\ref{eq:g3}), it is possible to simplify (\ref{eq:fr_sepatrix}) to see the roots clearly as follows,
\begin{equation}\label{eq:modifiedC}
\mathcal{C}(h_*) = (h^2_*-1) \left( h^2_* + \frac{1}{2} + \frac{1}{2}\sqrt{1 + \frac{4 \bar{g}_0}{6 \bar{g}_1 - \bar{g}_0}} \right)
\left( h^2_* + \frac{1}{2} - \frac{1}{2}\sqrt{1 + \frac{4 \bar{g}_0}{6 \bar{g}_1 - \bar{g}_0}} \right) =0.
\end{equation}

When $g_3 <0$, i.e. $\bar{g}_0 < 6 \bar{g}_1$, still there is only one (real) solution, $h_*=1$,
since $\sqrt{1 + \frac{4 \bar{g}_0}{6 \bar{g}_1 - \bar{g}_0}}$ must be imaginary.
Figure~\ref{fig:negativeg3} (Left) shows that we can find a proper inflationary trajectory around $h_*=1$ with $g_2>0$.
On the other hand, for $g_2<0$,
the direction of the trajectory is opposite when one crosses the line at which $h''$ diverges,
represented by the black dash in Figure \ref{fig:negativeg3} (Right).
As $h''$ is divergent, the terms multiplying $h''$ of (\ref{eq:frN_three}) become important.
On this divergence curve, these terms should vanish,
\begin{equation}\label{eq:h_diverge}
(-72 \bar{g}_2 h + 2592 g_3 h^3) + 1296 g_3 h h' = 0 \qquad \rightarrow \qquad h' = -2 h^2 + \frac{\bar{g}_2}{18 g_3} \, .
\end{equation}
The direction of the trajectory at very large $h$ is outward, e.g.\ both $h$ and $h'$ increase,
but since the divergence curve is \emph{over} $h_* = 1$,
the direction of the trajectory at $h \gtrsim 1$ is inward, e.g.\ $h$ decreases,
and also the direction at $h \lesssim 1$ is outward.
In the case $g_2 < 0$, therefore,
$h_* = 1$ is an attractor so that all nearby trajectories cannot go to the non-inflationary region.

\begin{figure}[h!]
\centering
\includegraphics[height=0.3\textwidth]{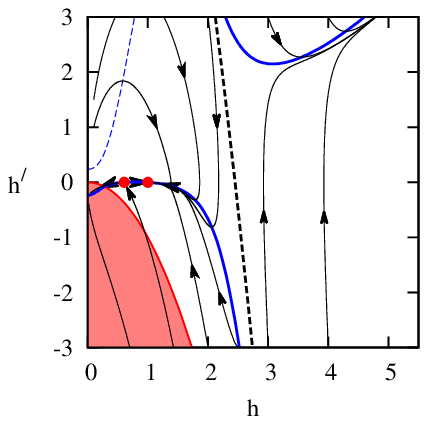}
\includegraphics[height=0.3\textwidth]{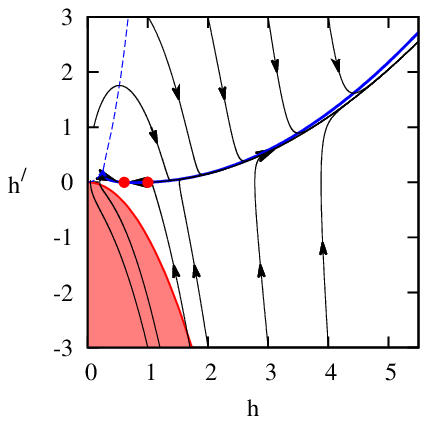}
\includegraphics[height=0.3\textwidth]{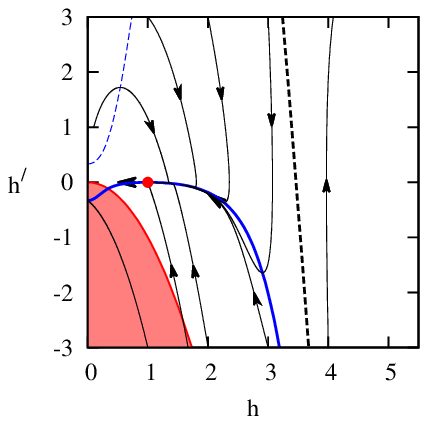}
\includegraphics[height=0.3\textwidth]{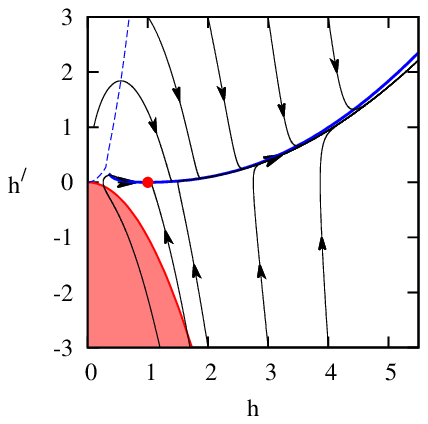}
\includegraphics[height=0.3\textwidth]{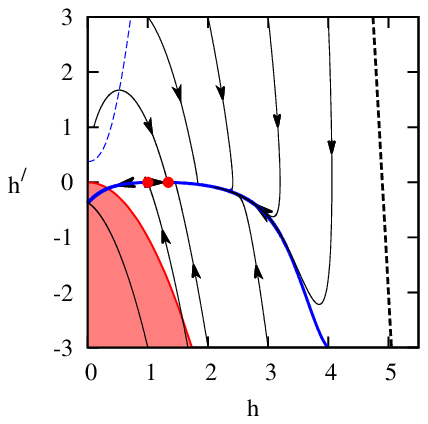}
\includegraphics[height=0.3\textwidth]{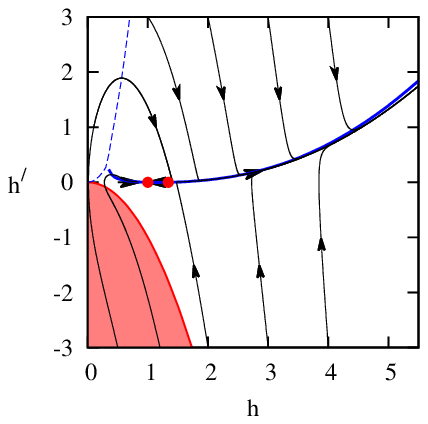}
\caption{
The general behavior of the classical trajectories in ($h, h'$) phase diagram with $g_3 > 0$.
Top: two roots with $h_{1*}=1$ and $h_{2*}<1$:
left: ($\bar{g_0}, \bar{g_1}, \bar{g_{2}},  g_{3}$) = (2, 1, 1, 1/216) and $N_\mathrm{efold} = 60$;
right: ($\bar{g_0}, \bar{g_1}, \bar{g_{2}},  g_{3}$) = (2, 1, -1, 1/216).
Middle: one with $h_{*}=1$:
left: ($\bar{g_0}, \bar{g_1}, \bar{g_{2}},  g_{3}$) = (4, 1, 1, 1/432) and $N_\mathrm{efold} = 100$;
right: ($\bar{g_0}, \bar{g_1}, \bar{g_{2}},  g_{3}$) = (4, 1, -1, 1/432).
Bottom: two roots with $h_{1*}=1$ and $h_{2*}>1$:
left: ($\bar{g_0}, \bar{g_1}, \bar{g_{2}},  g_{3}$) = (5, 1, 1, 1/864) and $N_\mathrm{efold} = 200$;
right: ($\bar{g_0}, \bar{g_1}, \bar{g_{2}},  g_{3}$) = (5, 1, -1, 1/864).
Red dot: the $h_*$-point. Red area: the non-inflationary era.
Blue solid: the asymptotic path(AP) where $h''=0$.
Blue dash: $h''=0$.
Black dash: the divergence curve where $h''$ diverges.}
\label{fig:positiveg3}
\end{figure}

When $g_3 > 0$, i.e. $\bar{g}_0 > 6 \bar{g}_1$, we have more possibilities.
From (\ref{eq:modifiedC}), there would be two real and positive roots,
\begin{equation}\label{eq:tworoots}
h_{1*} =1,~~~ \textrm{or}~~~ h= h_{2*} \equiv \sqrt{-\frac{1}{2}+  \frac{1}{2}\sqrt{1 + \frac{4 \bar{g}_0}{6 \bar{g}_1 - \bar{g}_0}} }.
\end{equation}
By straightforward calculations, one observes that there are three distinct cases,

\begin{itemize}
\item $\bar{g}_0 > 4 \bar{g}_1$:
$h_{1*} = 1$ and $h_{2*} > 1$.

If $g_2 > 0$, then from (\ref{eq:h_diverge}) the divergence curve is \emph{over} $h_*$'s,
and $h_{2*} > 1$ is an attractor and $h_{1*} = 1$ is a repeller.
Therefore, we can find a proper inflationary trajectory around $h_{1*} = 1$ (Top-left of Figure~\ref{fig:positiveg3}).

On the other hand, if $g_2 < 0$, then the divergence curve is \emph{under} $h_*$'s,
now $h_{2*} > 1$ is a repeller and $h_{1*} = 1$ is an attractor,
and there is no proper inflationary trajectory (Top-right of Figure~\ref{fig:positiveg3}).

\item $\bar{g}_0 = 4 \bar{g}_1$:
 $h_{1*} = h_{2*} = 1$.

 In this case, if $g_2 > 0$, then $h_* = 1$ is an attractor in the direction $h > h_*$
 and a repeller in the direction $h < h_*$,
 so we can find a proper inflationary trajectory around $h_* = 1$ (Middle-left of Figure~\ref{fig:positiveg3}).

 On the other hand, if $g_2 < 0$, then $h_* = 1$ is a repeller in the direction $h > h_*$
 and an attractor in the direction $h < h_*$,
 so there is no proper inflationary trajectory (Middle-right of Figure~\ref{fig:positiveg3}).

\item $\bar{g}_0 < 4 \bar{g}_1$:
 $h_{1*} = 1$ and $h_{2*} < 1$.
 This is the same as the case $\bar{g}_0 > 4 \bar{g}_1$,
 except that we can find a proper inflationary trajectory around $h_{2*} < 1$ (Bottom of Figure~\ref{fig:positiveg3}).
\end{itemize}

\subsubsection{e-foldings}
Assuming that there exists a trajectory passing near the $h_*$-point,
its e-foldings will come mainly from the region near the point.
Since after the trajectory passes the $h_*$-point as it follows the AP,
its position ($h_*+\Delta h, \Delta h'$) satisfies
\begin{equation}
\Delta h' = \frac{432 g_{3} h_* ^4 - \bar{g}_1}{18 h_* (36 g_{3} h_*^2 - \bar{g}_{2})} \Delta h \, ,
\end{equation}
which follows from (\ref{eq:frN_three}).
Therefore, the time scale to enter and escape the $h_*$-point is
\begin{equation}
\Delta t \sim 2 \frac{1}{H_0} \frac{\Delta h}{\Delta h'} =
\frac{1}{H_0} \frac{36 h_* (36 g_{3} h_* ^2 - \bar{g}_{2})}{432 g_{3} h_*^4 - \bar{g}_1} \, ,
\end{equation}
and the e-foldings generated near the $h_*$-point are
\begin{equation}\label{eq:fr_efold}
\Delta N^*_\mathrm{efold}  \sim \fn{H}{h_*} \Delta t =  \frac{36 h_* ^2 (36 g_{3} h_* ^2 - \bar{g}_{2})}{432 g_{3} h_*^4 - \bar{g}_1}  \gtrsim 60\, .
\end{equation}

If $g_3 = 0$, then $h_* = 1$ and (\ref{eq:fr_efold}) becomes
\begin{equation}\label{eq:fr_efold_g2}
36 \frac{\bar{g}_2}{\bar{g}_1} \gtrsim 60  \qquad \rightarrow \qquad
\frac{\bar{g}_2}{\bar{g}_1} \gtrsim \frac{5}{3} \, ,
\end{equation}
which is essentially the same result as (\ref{eq:efold1}).

If $g_3 \neq 0 $, then (\ref{eq:fr_efold}) becomes
\begin{equation}\label{eq:fr_efold_g2g3}
\frac{36 h_* ^2 (36 g_{3} h_* ^2 - \bar{g}_{2})}{432 g_{3} h_*^4 - \bar{g}_1}  \gtrsim 60 \qquad \rightarrow \qquad
3 \bar{g}_2 h_*^2 \gtrsim 5 \bar{g}_1 - 2052 g_3 h_*^4\, ,
\end{equation}
which means that if $\bar{g}_2$ is not too small then the classical trajectory can give enough e-foldings.

\subsection{General cases with higher derivative terms}
We can now extend our action by adding contractions of the Riemann tensor as
\begin{multline}\label{eq:general_action}
I_{\Lambda}[g] = -\int \d{^4 x} \sqrt{-g} \left[ \Lambda^4 g_0 + \Lambda^2 g_1 R
+ g_{2} R^2 + g_{2a} R^{\mu \nu} R_{\mu \nu} + g_{2b} R^{\mu \nu \rho \sigma} R_{\mu \nu \rho \sigma} \right. \\
\left. \Lambda^{-2} g_{3} R^3 + \Lambda^{-2} g_{3a} R R^{\mu \nu} R_{\mu \nu} + \Lambda^{-2} g_{3b} R R^{\mu \nu \rho \sigma} R_{\mu \nu \rho \sigma} \right] \, ,
\end{multline}
from which follows the relevant dynamical equation
\begin{equation}\begin{split}
\fn{\bar{\mathcal{N}}}{h,h',h''} &\equiv -\frac{2 \Lambda^2}{H_0 ^6} \left( \frac{\delta I_{\Lambda}}{\delta g_{00}}\right)_\mathrm{FRW} = 0 \\
\label{eq:general_nbar}
& = \left[ -\bar{g}_0 + 6 \bar{g}_1 h^2 - \left(864 g_{3} + 216 g_{3a} + 144 g_{3b} \right) h^6 \right] \\
& \quad + \left[ - \left(216 \bar{g}_{2} + 72 \bar{g}_{2a} + 72 \bar{g}_{2b} \right) h^2 + \left(7776 g_{3} + 2160 g_{3a} + 576 g_{3b} \right) h^4 \right] h' \\
& \quad + \left[ \left(36 \bar{g}_{2} + 12 \bar{g}_{2a} + 12 \bar{g}_{2b}\right) + \left(3240 g_{3} + 1008 g_{3a} - 216 g_{3b}\right) h^2 \right] h'^2  \\
& \quad - \left(432 g_{3} + 144 g_{3a} + 144 g_{3b}\right) h'^3 \\
& \quad  + \left\{ \left[ -\left(72 \bar{g}_{2} + 24 \bar{g}_{2a} + 24 \bar{g}_{2b}\right) h + \left(2592 g_{3} + 720 g_{3a}+ 288 g_{3b} \right) h^3 \right] \right. \\
& \quad \quad + \left. \left. \left(1296 g_{3} +432 g_{3a}+ 144 g_{3b} \right) h h' \right. \right\} h'' = 0 \, .
\end{split}\end{equation}

By comparing (\ref{eq:frN_three}) and (\ref{eq:general_nbar}),
one can see that the classical behavior of the quadratic terms in (\ref{eq:general_action})
can be reduced into that of the $\fn{f}{R}$ case simply by replacing $g_{2}$ to
\begin{equation}\label{eq:g2}
g_2 \qquad \rightarrow \qquad g_{2} + \frac{g_{2a} + g_{2b}}{3} \, .
\end{equation}
This fact can also be checked by considering the relations among the involved tensors and their special properties
that appear in four dimensions.
In four dimensions, the square of the Weyl tensor, $C_{\mu \nu \rho \sigma}$ is given as
\begin{equation}
C^2_{\mu \nu \rho \sigma}  =  R^2_{\mu \nu \rho \sigma} -2 R^2_{\mu \nu} +\frac{1}{3}R^2.
\end{equation}
The Euler scalar is given as
\begin{equation}
\chi_{_{\mathrm{Euler}}}  =  R^2_{\mu \nu \rho \sigma} -4 R^2_{\mu \nu} + R^2.
\end{equation}
The variation of the general quadratic action can be written as
\begin{eqnarray}
&&\delta  \int \d^4x \sqrt{-g} \left( g_2 R^2 + g_{2a} R_{\mu \nu}^2 + g_{2b} R_{\mu \nu \rho \sigma}^2 \right)  \nonumber \\
&&=
\delta \int \d^4x \sqrt{-g} g_{2} R^2 + \delta \int d^4x \sqrt{-g} g_{2a} \left( \frac{1}{2}C_{\mu \nu \rho \sigma}^2 -\frac{1}{2} \chi_{_{\mathrm{Euler}}} + \frac{1}{3}R^2 \right) \nonumber \\
&& + \delta \int \d^4x \sqrt{-g} g_{2b}\left( 2 C_{\mu \nu \rho \sigma}^2 -2 \chi_{_{\mathrm{Euler}}} + \frac{1}{3}R^2 \right).
\end{eqnarray}
The variation of the $C^2$ term around a conformally flat metric like FRW is zero.
The variation of Euler number vanishes identically (since it is a topological number of the manifold).
Around the FRW background, thus, one has
\begin{eqnarray}
\delta_{\textrm{FRW}} \int \d^4x \sqrt{-g} \left( g_2 R^2 + g_{2a} R_{\mu \nu}^2 + g_{2b} R_{\mu \nu \rho \sigma}^2 \right) \nonumber \\
= \left( g_2 + \frac{g_{2a}}{3} + \frac{g_{2b}}{3} \right) \delta_{\textrm{FRW}} \int \d^4x \sqrt{-g} R^2,
\end{eqnarray}
confirming (\ref{eq:g2}).

On the other hand, for the cubic order, we do not have the exact equivalence between (\ref{eq:general_action}) and $\fn{f}{R}$ at the classical level.
However, we can find a similar relation among $g_3$'s, which is crucial to determine the asymptotic behavior of the phase diagram,
by comparing $\fn{\mathcal{C}}{h}$'s.
From (\ref{eq:general_nbar}), we read off $\fn{\mathcal{C}}{h}$,
\begin{equation}
\fn{\mathcal{C}}{h} = 864 \left( g_{3} + \frac{g_{3a}}{4} + \frac{g_{3b}}{6}  \right) h^6  - 6 \bar{g}_1 h^2 + \bar{g}_0 = 0 \, .
\end{equation}
Then what we have discussed in the case of $\fn{f}{R}$ is still valid in the general case with (\ref{eq:general_action})
by replacing $g_3$ to
\begin{equation}
g_3 \qquad \rightarrow \qquad g_{3} + \frac{g_{3a}}{4} + \frac{g_{3b}}{6} \, .
\end{equation}
Once the derivatives of the Ricci scalar, Ricci and Riemann tensor and their powers are added to the Lagrangian, the general behavior will change.
However, this is beyond the scope of this work.

\section{Discussion}
\label{sec:discuss}

In this paper, we discussed the possibility of inflation on higher derivative theories in the context of asymptotically safe gravity.
Though the coupling constants are expected to be determined by the renormalization group flow,
it gives nontrivial constraints on them that our universe has experienced the inflationary era.
We could find the parametric relations between couplings by exploring successful inflationary trajectories on the phase diagrams in the bottom-up fashion.
The asymptotic behaviors around the $h_*$-points, which are slow-roll points, are crucial for generating enough e-foldings
and then escaping from the de Sitter state.

There are several topics for future works,
which may complete the map between higher derivative gravities and Einstein's gravity plus some scalar field theories.
First, as we include general higher derivative terms, the types and the number of the fields will increase and the nontrivial couplings between such fields will be introduced \cite{hindawi}.
Second, as in the standard Einstein side there are speculative themes such as eternal inflation and multiverse, it would be also intriguing to look for their counterparts in higher derivative gravities.

Moreover, this work is only the first step to a realistic cosmological scenario from asymptotically safe gravity, 
and it remains how to complete the whole cosmic history with being consistent with cosmological observations.
We are hoping that higher derivative terms are initially dominant to produce the inflationary phase in the high energy scale, 
and then allow Einstein's gravity in the low energy scale.
To have such a consistent cosmic history, it would be crucial to get a nontrivial evolution of the coupling constants.
One may find possible ways by considering the dilatonic dependence on the couplings from string theory (e.g. \cite{dilatonic})
or the energy scale dependence from the renormalization group flow \cite{renormalization}.

Also, in order to fit the asymptotically safe gravity into the observation, 
we need to study the cosmic perturbations during inflation and their evolution after inflation.
To do this properly, we need to understand how to have proper scalar, vector and tensor modes of the metric perturbations and make them evolve to produce the correct observations such as the power spectrum, the CMB and the production of gravitational waves.
One may get some hints from previous studies about the cosmological perturbations, gravitational waves and cosmological vorticity within the quadratic theory \cite{hwang}.
Finally, since the evolution of perturbations depends on the matter contents after reheating, it is also crucial to look for a reheating mechanism.

\section*{Acknowledgements}
The authors thank to Bayram Tekin, Ewan Stewart, Hassan Firouzjahi, Wan-il Park, Tahsin Sisman, Dong-han Yeom,  Alexei Starobinsky,
Vincent Vennin, Chang Sub Shin, Dong-il Hwang and Roberto Percacci.
HZ thanks the hospitality of IPM where this work was progressed substantially.
The authors are supported by Basic Science Research Program through the National Research Foundation of Korea(NRF)
funded by the Ministry of Education, Science and Technology(2009-0077503).
SEH is also supported by the National Research Foundation
grant (2009-006814, 2007-0093860) funded by the Korean government.
HZ is also supported by T\"{U}B\.{I}TAK research fellowship programme for foreign citizens.

\end{document}